\documentclass[jcp,twocolumn,groupedaddress,reprint,superscriptaddress,showpacs,preprintnumbers,amsmath,amssymb,aps, mathtools]{revtex4-1}
\usepackage{textcomp}
\usepackage{graphicx}% Include figure files
\usepackage{dcolumn}% Align table columns on decimal point
\usepackage{bm}%
\usepackage[colorlinks=true,citecolor=blue]{hyperref}
\hypersetup{colorlinks=true,citecolor=blue,linkcolor=red,urlcolor=blue}
\newcommand {\be}{\begin{equation}}
\newcommand {\ee}{\end{equation}}

\usepackage{ulem}
\usepackage{color}

\begin{document}
	
	\title{High-harmonics and isolated attosecond pulses from MgO}
	
	\author{Zahra Nourbakhsh}
	\email[]{zahra.nourbakhsh@mpsd.mpg.de}
	\affiliation{Max Planck Institute for the Structure and Dynamics of Matter, Luruper Chaussee 149, 22761 Hamburg, Germany.}
	
	\author{Nicolas Tancogne-Dejean}
	\email[]{nicolas.tancogne-dejean@mpsd.mpg.de}
	\affiliation{Max Planck Institute for the Structure and Dynamics of Matter, Luruper Chaussee 149, 22761 Hamburg, Germany.}
	
	\author{Hamed Merdji}
	\affiliation{LIDYL, CEA, CNRS, Université Paris-Saclay,CEA Saclay 91191 Gif sur Yvette, France.}
	
	\author{Angel Rubio}
	\email[]{angel.rubio@mpsd.mpg.de}
	\affiliation{Max Planck Institute for the Structure and Dynamics of Matter, Luruper Chaussee 149, 22761 Hamburg, Germany.}
	%\affiliation{Center for Free-Electron Laser Science CFEL, Deutsches Elektronen-Synchrotron DESY, Notkestraße 85, 22607 Hamburg, Germany.}
	\affiliation{Nano-Bio Spectroscopy Group and ETSF, Departamento de Fisica de Materiales, Universidad del Pa\'is Vasco UPV/EHU, 20018, San Sebasti\'an, Spain.}
	\affiliation{Center for Computational Quantum Physics (CCQ), The Flatiron Institute, 162 Fifth Avenue, New York, New York 10010, USA.}
	
	\begin{abstract}
		On the basis of real-time {\it ab initio}  calculations, we study the non-perturbative interaction of two-color laser pulses with MgO crystal  in the strong field regime to generate isolated attosecond pulse from high-harmonic emissions from MgO crystal.
		In this regard, we examine  the impact of incident pulse characteristics such as its shape, intensity, and ellipticity as well as the consequence of the crystal anisotropy on the emitted harmonics and their corresponding isolated attosecond pulses.
		Our calculations predict the creation of isolated attosecond pulses with a duration of $\sim$ 300 attoseconds; 
		in addition, using elliptical driving pulses, the generation of elliptical isolated attosecond pulses is shown.
		Our work prepares the path for all solid-state compact optical devices offering perspectives beyond traditional isolated attosecond pulse emitted from atoms.
	\end{abstract}

	\maketitle

\section{Introduction}\label{intro}

High-harmonics generation (HHG) is a nonlinear optical process as the result of a strong laser field interacting with either an atom, a molecule, a plasma or a crystal.
In this phenomenon, the target system emits light at frequencies equal to the integer multiples of the frequency of the driving laser in the classical multicycle regime. 
HHG was discovered in 1987 in gases and atomic systems \cite{ghhg1, ghhg2}; it has been extended to solids and condensed matter systems since 2010 \cite{hhgsol}. Today, after more than three decades, HHG is an alive and promising research area from both fundamental and practical points of view. 
HHG is a source of coherent extreme ultraviolet (EUV) radiation that has pioneered numerous applications. 
In particular, HHG in atoms has established a new area of research so-called attosecond science. 
The creation of an isolated attosecond pulse (IAP) from HHG light source in 2001 \cite{ashhg} was a milestone in the HHG history and ultrafast technology. After that, continuous efforts have been produced to generate brighter and shorter IAPs from the coherent HHG sources \cite{as}. IAPs can reveal microscopic details of the physical processes involved, with attosecond timescale resolution, such as electron motion in materials, bond creation or bond breaking, and ultrafast, sub-optical-cycle,  quantum-mechanical phenomena \cite{as1, as2, as3, caliap}.

Solid-state HHG has attracted a lot of attention in the last years
\cite{rev2019,hhgsol1,hhgsol2,hhgsol3,hhgsol4,hhgsol5,hhgsol7,hhgsolT1,hhgsolT2,hhgsolT3,mgo,mgo-aniso,mgo-anis2,mos2,sio2,intint1,intint2,intint3,vdw,el0}
due to the new possibilities in solids to control and taylor the HHG radiation properties. 
It is possible act on the HHG emission by engineering the crystal structure, chemically or mechanically \cite{hhgsol1,hhgsol2,hhgsol3,hhgsol4}, or rotating the target sample \cite{ mgo-anis2,mgo-aniso}. Additionally, since the HHG mechanisms in solids are described by interband and intraband contributions \cite{mos2,intint1,intint2,intint3}, the coupling between electrons and holes in crystals makes it possible that the excited electrons 
recombine in neighboring sites \cite{el0}. Accordingly, solids could show stronger response to elliptical laser field in comparison with the gas phase where the excited electrons need to find their parents ion; note that ellipticity behaviors in solids depend on the material properties \cite{hhgsolT2,hhgsolT3,mgo}. Moreover, higher density \cite{hhgsolT1} and electronic momentum change induced by the lattice periodicity \cite{mgo} cause stronger HHG spectrum in solids; for instance, mono-atomic crystals display brighter HHG spectrum compare their gas phases under the same incident pulse \cite{vdw}. Furthermore, high-harmonics response could inform about the electronic and dynamical properties of the solid system \cite{rev2019}.

The shortest IAP ever produced was achieved in 2017 with a duration of  53 attoseconds (as) \cite{ias1}; following more than a decade of advances \cite{ias2, ias3, ias4}.Despite these successes, since the corresponding applied methods are difficult to implement, the attempts toward producing more accessible techniques are continued. In this regard, beside the input pulse properties, the target system  and its strong field response is very important. 

In this article, we demonstrate bright IAPs from a solid target. Controlling the driving pulse using two or more colors and tuning the second pulse parameters in order to confine the emission probability in a shorter time than its duration  was originally proposed in atoms and in a free electron laser \cite{tcp1,tcp2}.
For the target system, we consider wide bandgap (7.8~eV) MgO crystal. MgO is a well-known solid in HHG community due to its  high damage threshold against intense ultrashort infrared input pulses. In addition, because of ionic bonds in MgO crystal with rocksalt (NaCl) structure, its inhomogeneous electron-nuclei potential  is similar to atomic cases; consequently MgO crystal gathers the advantages \cite{hhgsolT1} of both atomic and solid systems in one material. 

Our demonstration is based on {\it ab initio} time-dependent density-functional theory (TDDFT) \cite{tddft1, tddft2} implemented in octopus package \cite{oct}  (see the Method section for more details). This method allows us to model the electron dynamics in the solids without making strong assumptions and was shown to provide an appropriate agreement between simulation results and experimental measurements \cite{hhgsolT2}. We will show that using two-color intense pulses, IAPs of duration as short as $\sim$ 300~as are extracted from the harmonic emission in EUV range. This is shorter than what was measured experimentally in SiO$_2$ nanofilm (470 as) \cite{sio2}, or the {\it ab initio} prediction of IAP duration in MoS$_2$ monolayer (2280 as) \cite{mos2}.

The paper is organized as follows. Having reviewed the theoretical methods in the next section, we present our results in Sec.~\ref{result}. The role of  pulse strength, polarization direction, and impact of two-color asymmetric pulses  on HHG response as well as IAP creation are discussed in this section. We end Sec.~\ref{result} with the investigation of ellipticity effect on HHG and IAP  production. Finally, Sec.~\ref{summary} summarizes our main results.

\section{Method}\label{method}

TDDFT calculations of the time evolution of the electronic wave functions are performed using the octopus code \cite{oct} on the basis of the  Kohn-Sham equation, defined as:
\be
\begin{split}
i \frac{\partial}{\partial t} \phi_i(\mathbf{r},t) = 
\Big(-\frac{\nabla^2}{2} \mathbf + v_{ext}(\mathbf{r},t) + v_H[n(\mathbf{r},t)] \\
+ v_{xc}[n(\mathbf{r},t)] \Big)  \phi_i(\mathbf r,t)\,,
\end{split}
\ee
in this equation $v_{ext}(\mathbf{r},t)$ is the external potential including the applied synthesized laser field and nuclear potentials, $v_H$ is Hartree part of the Coulomb electron-electron interaction,  $v_{xc}$ is exchange-correlation potential, $n(\mathbf{r},t)$ is the time dependent electron density defined as $n(\mathbf{r},t) = \sum_{i} |{\phi_i(\mathbf{r},t)}|^2$, with $\phi_i$ the Kohn-Sham orbital associated with the index $i$ corresponding to both a band and a \textbf{k}-point indexes. In the next step, the total microscopic current, $\mathbf{j}(\mathbf{r},t)$, is computed from the time-dependent wavefunctions. Finally, the high-harmonics spectrum is obtained by the Fourier transform of the laser driven electron current
\be
\label{hhgeq}
HHG(\omega)= \Big|FT \left(  \frac{\partial}{\partial t} \int_{\Omega}  d^3\mathbf{r}~ \mathbf{j}(\mathbf{r},t) \right) \Big|^2\,,
\ee
where $\Omega$ is the system volume. The attosecond pulse can be extracted from the coherent superposition of consecutive harmonics \cite{caliap} 
\be
I(t)= \Big|\sum_{\omega_i}^{\omega_f} e^{i\omega t} ~ FT \left(  \frac{\partial}{\partial t} \int_{\Omega}  d^3\mathbf{r}~ \mathbf j(\mathbf{r},t') \right) \Big|^2\,,
\ee
$\omega_i$ and $\omega_f$ define the energy window used to calculate the IAP.

Unless stated differently, the exchange-correlation term is described by local density approximation (LDA) and we used norm-conserving pseudopotentials.
While LDA underestimate the bandgap, it correctly described the band dispersion of the valence and conduction bands \cite{ldaimp}, and thus is capable of describing properly the coupled interband and intraband dynamics.
A dense k-point grid of $28\times28\times28$ and a grid spacing equal to 0.2 bohr are used thorough our calculations.

\section{Results and discussion}\label{result}
\subsection{Linear polarized pulse: high-harmonics and IAP generation}
\subsubsection{One-color versus two-color pulses HHG}

\begin{figure*}
	\centering
	\includegraphics[width=1.\linewidth]{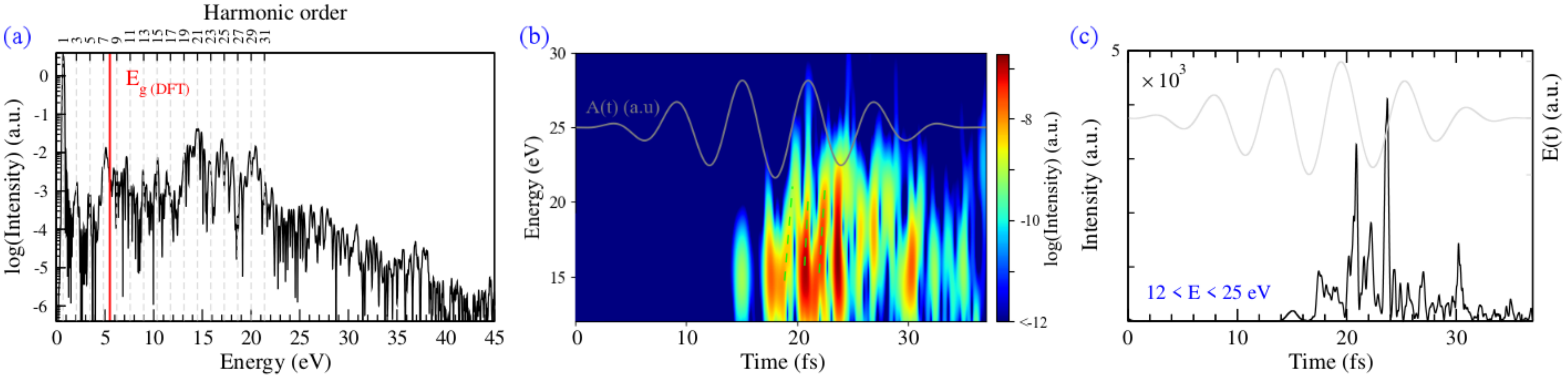}
	\caption{\label{single} 
		(a) HHG spectrum corresponding to the fundamental laser; the pulse polarization is along [100] direction. The dashed lines mark the harmonic frequencies. The solid red line indicates the MgO DFT bandgap (5.4~eV). 
		(b) Time-frequency analysis of the HHG; the time window to calculate the Gabor transform is taken to be 0.25~fs. This figure also shows the time profile of the vector potential.  The green arrows indicate the attosecond chirp.
		(c) A train of attosecond pulses resulted from the fundamental laser pulse for the given energy window. The time-dependent electric field is also displayed.}
\end{figure*}

We start our discussion by analyzing the non-perturbative interaction between our fundamental pulse, polarized along the [100] direction, 
and the MgO crystal; the results are presented in Fig.~\ref{single}.
Our fundamental laser pulse is set to a wavelength  of 1.8~$\mu$m.
The intensity of few $10^{12}$  W/cm$^2$, corresponding to an electric field strength of $\sim$ 0.6~V/\AA~ in the bulk system, is kept below the damage threshold of MgO crystal. The pulse duration at full width at half-maximum (FWHM) is equal to 18~fs with a $\sin^2$ envelope shape for the vector potential; the carrier-envelope phase of the fundamental pulse is set to zero.

The HHG spectrum, presented in Fig.~\ref{single}a, shows that odd harmonics up to cutoff energy of $\sim$ 22~eV are generated while, because of the inversion symmetry in MgO crystal, even harmonics are absent here \cite{intint3}.

The mechanism of solid state HHG is usually explained by interband and intraband dynamics \cite{hhgsol2,mos2}. The first four harmonics in Fig.~\ref{single}a, with energies below the DFT bandgap, are generated by intraband acceleration while for harmonics above the bandgap, interband recombination can also contribute to the emission of harmonics. As a result, the spectrum above the bandgap is more noisy and not as clean as the HHG signal originating from intraband dynamics only. This behavior is discussed in more details in Ref.~\cite{hhgsolT1} for bulk silicon. 

The time-frequency analysis, presented in Fig.~\ref{single}b, predicts that no IAPs could be extracted from this harmonic spectrum, 
and as Fig.~\ref{single}c shows, a train of attosecond pulses results from this emission. 
At lower energy, i.e. between 12 and 20 eV, the HHG emission is maximum and shows a complex dynamics with evidence of a positive chirp of the attosecond bursts (see green dashed lines).

\begin{figure*}
	\centering
	\includegraphics[width=1\linewidth]{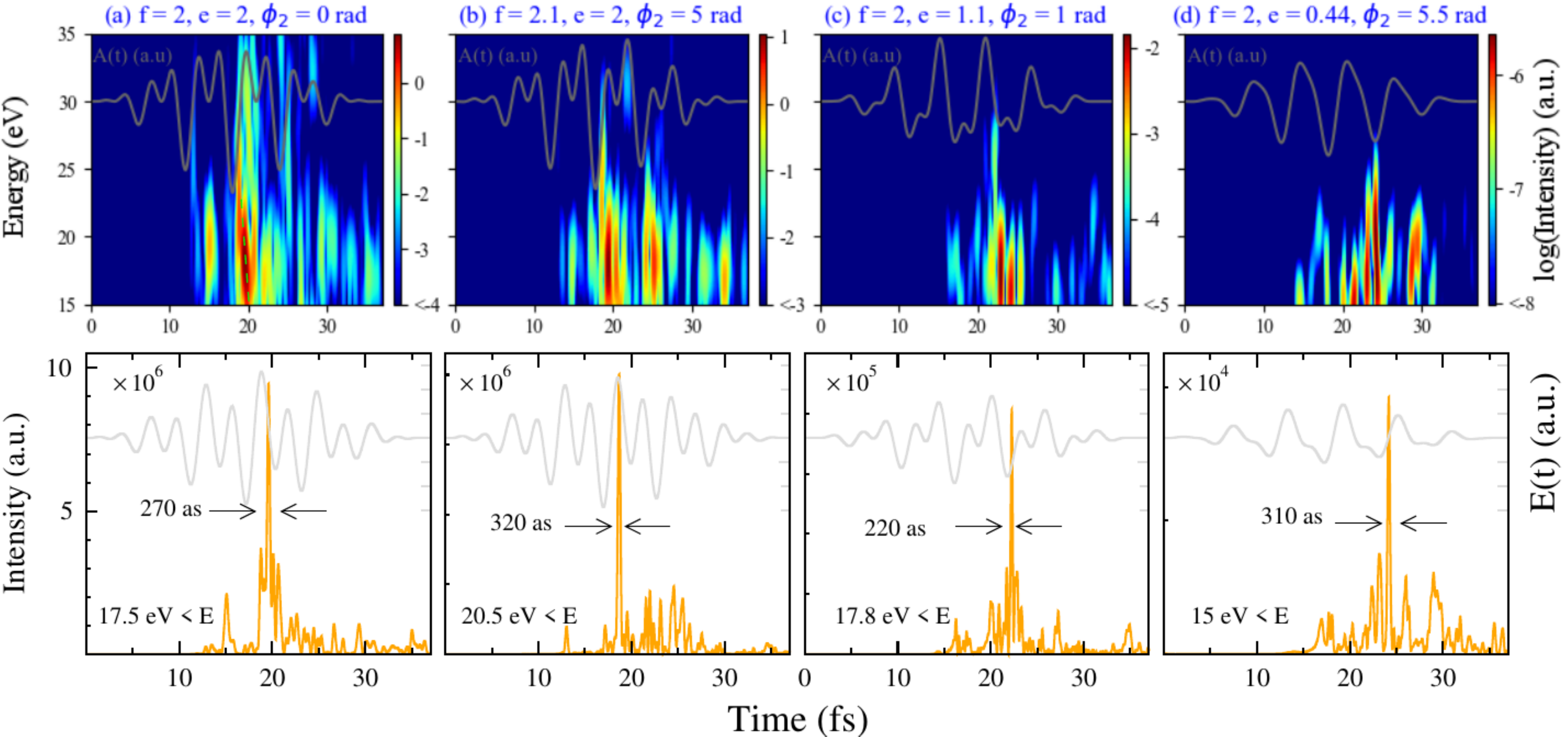} 
	\caption{\label{IAPs}  Two-color pulse time-frequency analysis (top row) and their related IAPs (bottom row). 
		The results for four different two-color pulses are shown; the light polarization is along [100] direction. The titles specify the pulse characteristics; e, f and $\phi_2$ as defined in Eq.~\ref{ef}. The green arrow in panel (a) shows a negative chirp. The IAP duration and the corresponding energy windows are given in the bottom panels. The minor spaces in IAP diagrams are equal to 1~a.u. and their scales are marked next to the axes. The total electric field profiles are also shown in light gray in the bottom row, using the same scale for the panels.}
\end{figure*}

The two-color results are illustrated in Fig.~\ref{IAPs}. The corresponding HHG spectra are displayed in Supplementary Fig.~S2.
As Eq.~\ref{ef} displays, the pulse shape and HHG spectra are controlled using the second-color pulse strength, frequency, and phase.
Practically, these characteristics are adjusted to control the attosecond electron dynamics within the MgO bands such that the HHG is emitted only during a fraction of an optical cycle over the entire pulse. 
The driving two-color vector potential is defined as 
\be
\label{ef}
\mathbf{A}(t) = \mathbf{A}_0 \sin^2 (t\pi/\delta) [\sin(\omega_0 t) + e/f \sin (f\omega_0 t + \phi_2)],
\ee
where $\mathbf{A}_0$ and $\omega_0$ are respectively the vector potential and frequency of the fundamental pulse,
$t$ is time, and the width $\delta$ is 36~fs consisting of twice the pulse duration. The pulse polarization, duration, and the envelope shape of the second laser field are the same as those of the fundamental pulse.
IAP are strongly governed by the effective detuning and dephasing of the second harmonic as shown in Fig.~\ref{IAPs}.
This reveals the importance of asymmetric pulse in IAP generation.

In Fig.~\ref{IAPs}, the total electric field strength increases from right 
to left; the peak field strength in Figs.~\ref{IAPs}d and \ref{IAPs}a are respectively 0.8~V/\AA~ and 1.7~V/\AA.~ 
This enhancement rises the emission intensity and extends the HHG cutoff energy from $\sim$ 22~eV in Fig.~\ref{IAPs}d to $\sim$ 42~eV in Fig.~\ref{IAPs}a (See Supplementary Figs.~S2 and S5. 
Practically, brighter IAP are produced slightly after the maximum electric field gradient which corresponds to the strongest acceleration of the electron current in the conduction band.
As indicated in Fig.~\ref{IAPs}, for each case according to its time-frequency spectrum, we reoptimized the energy window to obtain the shortest IAP possible from the harmonic emission, as one would do experimentally. 
Interestingly, the IAP duration which is reported in Fig.~\ref{IAPs} (defined from the FWHM of the filtered signal) do not show any serious dependence to electric field strength.

The IAP in Fig.~\ref{IAPs}a has a negative attochirp for the harmonics between 15-21~eV.
The negative chirp sign is opposite to the single color case and to the chirp usually reported in gas HHG \cite{chirp2}. Besides the material, the pulse properties like its wavelength, strength, and shape affect the chirp sign and magnitude. 
The fact that we can taylor the attochirp under certain driving conditions opens new perspectives for the control of attosecond pulses. Additionally, this could reveal details on the microscopic electron dynamics at place, but a detailed study goes beyond the scope of the present study.

Since DFT calculations performed at the level of the LDA or GGA (generalized gradient approximation) underestimate the bandgap,
we also performed calculations using the TB09 meta-GGA functional \cite{mgga} which yield an accurate estimation of the MgO bandgap. The results presented in Supplementary Fig.~S1 imply that the gap enhancement does not affect significantly our main conclusions, and shows the validity of using the LDA to study the generation of IAPs in MgO.

Our {\it ab initio} results for the IAP duration in MgO is shorter than what was obtained from TD-DFT calculations in MoS$_2$ crystal with FWHM duration of 2280~as \cite{mos2}. This is also shorter than the IAP measured experimentally in SiO$_2$ nanofilms with the duration of 470~as resulting from a single-cycle pulse with peak field strength of 1.1~V/$\AA$.  
Moreover, the energy windows used to extract the IAPs, which are specified in the bottom row of Fig.~\ref{IAPs}, are much wider than the calculated bandwidth in MoS$_2$ (16-20~eV) \cite{mos2} and that measured in SiO$_2$ (18-28~eV) \cite{sio2}. For example, the filtering area in Fig.~\ref{IAPs}a starts from 17.5~eV and cover the extreme-ultraviolet (EUV) energy range up to 
43~eV. Our results therefore show that MgO a promising candidate for the generation of IAPs for solid targets. Finally, we note that a recent publication \cite{mgo-as} on bulk MgO, using the one-dimensional semiconductor Bloch equation, predicts IAPs with a duration of 817-1000~as and a bandwidth of 20-35~eV for a two-cycle pulse with the wavelength of 1.6~$\mu$m.

Another worth mentioning point is the emission time of the intense attosecond pulses versus the electric field and vector potential extrema. For instance, in Fig.~\ref{IAPs}a, the IAP is emitted at the vector potential peak while  in Fig.~\ref{IAPs}b, the IAP is emitted at the electric field extremum; similar behaviors are seen in Fig.~\ref{single} or Supplementary Fig.~S4. Note that since within the dipole approximation $\mathbf{E}(t) \propto \frac{\partial}{\partial t} \mathbf{A}(t)$, the vector potential extrema correspond to zeros of the electric field. This could arise from the different microscopic origins of the harmonics involved in each IAP. For instance, the intraband dynamics is a coherent emission occurring in phase with the driving electric field, and is enhanced when the band structure has its largest curvature or when the electric field is in its maximum \cite{hhgsol5};
on the contrary, when the vector potential is maximum, the bandgap increases and the emitted harmonics are dominated by interband dynamics\cite{hhgsol5}.

We end this part by discussing the electric field strength and wavelength dependence of HHG cutoff energy. 
In gases, this dependency is defined by the ponderomotive energy $U_{\mathrm{p}} \propto \lambda^2 E^2_0$, but since the excited electrons in solids are not free  particles in a continuum but quasi-free particles moving on energy bands, this equation is not generally valid for solids. As shown in Supplementary Fig.~S5, we found that the harmonics cutoff scales linearly with the driving pulse peak field. 
However, besides the electric field strength, the pulse shape affects the cutoff energy, and we found that asymmetric pulses yield shorter energy cutoffs than single-color pules. 
According to our results, cutoff energy is found to be independent from pulse wavelength in MgO.
Note that opposite reports have been published about sensitivity \cite{hhgsol2, wldep1,wldep2} or non-sensitivity \cite{hhgsol1, hhgsolT1,wlind} of cutoff energy to wavelength.

\subsubsection{Anisotropy impacts}\label{anis}

\begin{figure*}
	\centering
	\includegraphics[width=0.9\linewidth]{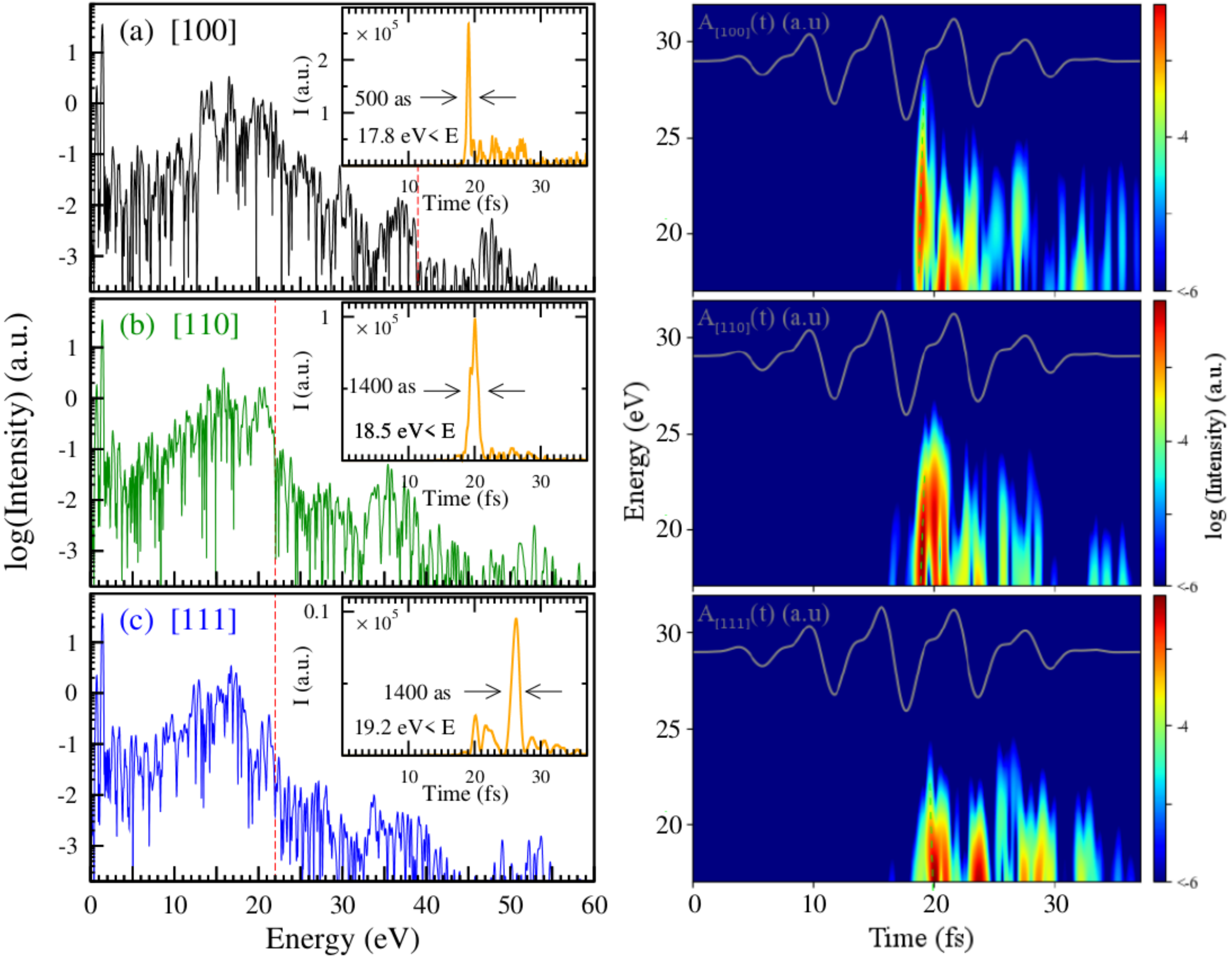}	
	\caption{\label{aniso}  The impact of pulse polarization direction on high-harmonics and IAP generation.
		Left panels: The pulse polarizations are respectively along [100], [110], and [111] from top to bottom. The dashed red lines specify the HHG cutoff energies.  The inset plots perform the corresponding IAPs. Right panels: The corresponding time-frequency analysis. The time window to calculate the Gabor transform is taken to be 0.25~fs. This figure also shows the time profile of the vector potential. The green arrows allow following the chirp of the pulse.}
\end{figure*}

The rest of our study on linearly polarized pulse in MgO focuses on the HHG response to the crystal anisotropy. Rotating the crystalline target or pulse polarization direction provides a way to manipulate the crystal harmonics emission which is not accessible to atoms, as solids possess intrinsic symmetries which influence the dipole coupling strength and possible transitions between bands, thus changing interband and intraband dynamics, and consequently, affecting the harmonics spectra. 

Figure~\ref{aniso} shows MgO HHG spectra as well as the time-frequency profiles arising from the same driving pulses but with different polarization directions; for the sake of simplicity, we just consider the light polarization along the high symmetry [100], [110], and [111] directions.
A two-color pulse is used to perform these calculations with f = 2, e = 0.9 and $\phi_2 = 1$ rad and the same polarization direction.

Regarding Fig.~\ref{aniso}, the harmonics emission, specially for energies above 20~eV,  
are suppressed for the [110] and [111] directions. The much stronger high-harmonics emission, for polarization along the [100] direction, could be roughly explained by the strong ionic potential  along this direction (see supplementary information). 
The potential along [111] direction is also between Mg$^{2+}$ and O$^{2-}$ ions, but the  
distance between ions in this direction is almost twice the bond length in [100]. 
Figure~\ref{aniso} displays HHG spectra for the [110] and [111] directions have the same cutoff energies; however, the spectrum with polarization along [110] is relatively more intensive. 
We note that the anisotropic HHG emission discussed here is consistent with the previous studies on MgO \cite{mgo-aniso, mgo-anis2}. 

The insets plots in Fig.~\ref{aniso} show the corresponding IAPs.  The results imply that for the [110] and [111] polarized pulses the length and contrast  of IAP generation rapidly deteriorates compared to  the [100] polarized pulse. 
The [100] case shows the brightest IAP and the shortest pulse duration with a slight positive chirp. The [110] case demonstrates a nicely contrasted attosecond pulse, however, the time frequency representation shows that a complex dynamic occurs with a pronounced positive chirp. The [111] case does not exhibit a well contrasted IAP. The main pulse (around 26 fs timing) has a pre-pulse with negative chirp (starting at 20fs). Again, this illustrates the complex attosecond dynamics occurring within the MgO bands. We conclude that, in addition to the pulse asymmetry, the calculated IAPs are strongly influenced by the crystal orientation, which is an important difference compared to the atomic case.

\subsection{Ellipticity impacts on high-harmonics and IAP generation}

\begin{figure*}
		\centering
	\includegraphics[width=0.9\linewidth]{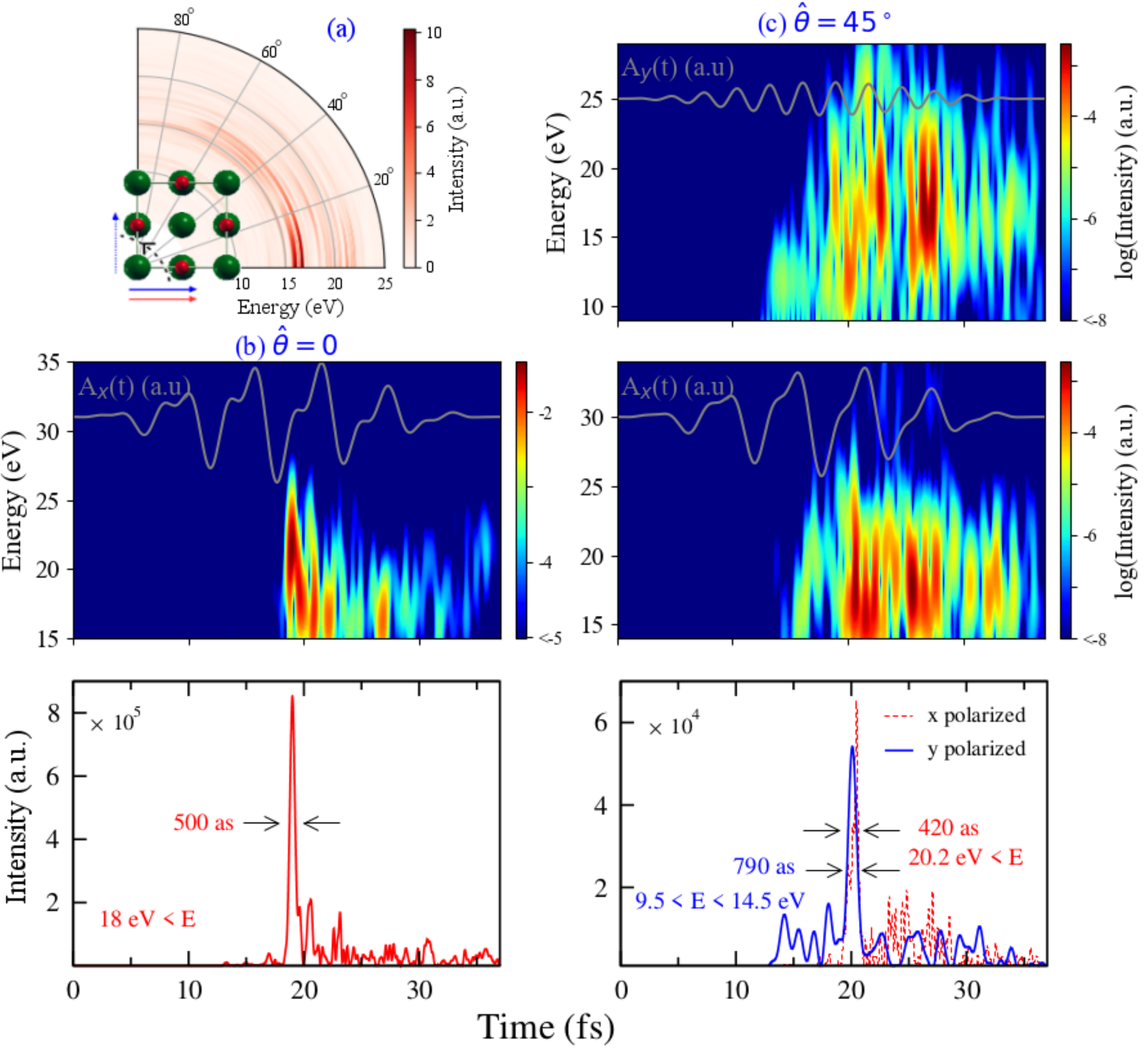}
	\caption{\label{asell}  Ellipticity impacts on high-harmonics and IAP generation.
		(a)  Two-color pulse HHG spectrum as a function of the second field polarization direction; the first field polarization is along [100] direction.
		(b, c) The polarization resolved calculated time-frequency spectra and IAPs:
		(b) for the non-rotated case or the linear polarized two-color pulse;
		(c) for the second field polarization rotation of $\theta = 45^{\circ}$;
		The first and the second rows show respectively the time-frequency plots for the $y$ and $x$-polarized harmonics; for the non-rotated case displayed in (b), high-harmonics are totally polarized along pulse polarization ($x$ direction here).
		The polarization resolved vector potential profiles are also shown, using the same scale for the panels.}
\end{figure*}

Using linearly-polarized two-color pulses allow generating customized elliptically-polarized pulses by rotating the polarization direction of one of the two pulses.
In order to investigate the ellipticity impact on HHG and IAPs in MgO, we performed calculations for non-collinear  polarized drivers.  
Figure~\ref{asell} summarizes our results for two-color laser pulses as a function of the second pulse polarization angle. 
The direction-resolved time-frequency profiles and their corresponding IAPs are shown in this figure.
Our driving vector potential is $\mathbf{A}(t) = A_0 \sin^2 (t\pi/\delta) [\sin(\omega_0 t)~ \hat{i} + \frac{1}{2.1} \sin (2.1\omega_0 t + 5.5)~ \hat{\theta} ]$ where $A_0$ and $\omega_0$ are respectively the fundamental pulse vector potential and frequency, as defined in Sec.~\ref{method}. 

The polar plot  in Fig.~\ref{asell}a shows that by rotating the second field, harmonics emissions continuously fall off; for instance, two intense harmonics with energies $\sim$ 16-17~eV drop approximately by an order of magnitude when the second pulse rotates from the parallel to the perpendicular direction. Besides, the cutoff energy decreases from 36~eV for collinear case to 27~eV and 25~eV for rotation angles equal to $45^{\circ}$ and $90^{\circ}$, respectively; more details are shown in the Supplementary Figs.~S6 and S7.

We now study the possibility of IAP generation for the second pulse rotation angles of 45$^\circ$ and  90$^\circ$ as well as the collinear case (or zero degree).
The time-frequency and the corresponding IAP plots for the high-harmonic emissions of the collinear and the rotation angle of 45$^\circ$ are respectively shown in Figs.~\ref{asell}b and \ref{asell}c;
for 90$^\circ$ rotation (not shown), no clear IAP is obtained from either $x$- or $y$-polarized high harmonics.  
Figure~\ref{asell}c reveals the possibility of having an elliptical IAP resulting from appropriate tuning of the different input parameters. 
As shown in this figure, the $x$ and $y$ polarized IAPs are emitted in the same time.
The elliptical IAPs are usually cumbersome to generate in gases and are important for studying spin polarized electronic motion in molecular or condensed-matter systems \cite{spinhhg1, spinhhg2}.

\section{Summary}\label{summary}
We study the nonlinear response of MgO crystal to incident intense pulses, we have employed  {\it ab initio} TDDFT calculations.
Using infrared asymmetric pulses with a duration of 18~fs and an intensity of $\sim~10^{13}$~W/cm$^2$, the following results have been obtained.

The generation of harmonics up to 43~eV, and IAPs as short as $\sim$~300~as have been predicted.  
Attochirp which is a feature of interband recombinant emissions \cite{chirp1}, can be manipulated from positive to negative and appears for the harmonics between 15-21~eV; at this energy range, there is a plateau in the HHG spectrum (for instance see Supplementary Fig.~S2), and the lying conduction bands with the Van-Hove singularities are presented in the band structure (see Supplementary Fig.~S3). 
MgO harmonic emission shows an anisotropic behavior, and HHG emission  for pulse polarization along  [100] direction is stronger. Furthermore, the efficiency of the generated IAP strongly decreases when the pulse polarization is not along [100] direction.
Finally, we have shown that the HHG signals drop rapidly for the elliptical polarized pulses; however, it provides the availability of easily generating elliptically polarized IAPs. 
The linear dependence of cutoff energy on the driving laser peak field has been observed. In addition, the effects of laser wavelength, its ellipticity, and the crystal anisotropy on cutoff energy have been discussed.

The results presented in this paper demonstrated the potential of solid-state materials in future ultrafast technologies.

\acknowledgments
This work has been supported by the European Research Council (ERC-2015-AdG694097), the Deutsche Forschungsgemeinschaft through the Priority Programme Quantum Dynamics in Tailored Intense Fields (QUTIF), Grupos Consolidados (IT1249-19), the Cluster of Excellence ’Advanced Imaging of Matter’ (AIM) and the Max Planck - New York City Center for Non-Equilibrium Quantum Phenomena.  The Flatiron Institute which is a division of the Simons Foundation is acknowledged.  
H.M. acknowledges support from the PETACom FET Open H2020 grant number 829153, OPTOLogic FET Open H2020 grant number 899794, DGA RAPID grant \textquotedblleft SWIM\textquotedblright and from the C\textquotesingle NANO research program through the NanoscopiX grant, and the LABEX \textquotedblleft PALM\textquotedblright (ANR-100LABX-0039-PALM) through the grants \textquotedblleft Plasmon-X\textquotedblright, \textquotedblleft STAMPS\textquotedblright and \textquotedblleft HILAC\textquotedblright. We acknowledge the financial support from the French ASTRE program through the \textquotedblleft NanoLight\textquotedblright  grant. The fruitful discussion with O. D. M\"ucke is acknowledged.

\title{ Supplementary Material: High-harmonics and isolated attosecond pulses from MgO}

\author{Zahra Nourbakhsh}
\affiliation{Max Planck Institute for the Structure and Dynamics of Matter, Luruper Chaussee 149, 22761 Hamburg, Germany.}
%\affiliation{Center for Free-Electron Laser Science CFEL, Deutsches Elektronen-Synchrotron DESY, Notkestraße 85, 22607 Hamburg, Germany.}

\author{Nicolas Tancogne-Dejean}
\affiliation{Max Planck Institute for the Structure and Dynamics of Matter, Luruper Chaussee 149, 22761 Hamburg, Germany.}
%\affiliation{Center for Free-Electron Laser Science CFEL, Deutsches Elektronen-Synchrotron DESY, Notkestraße 85, 22607 Hamburg, Germany.}

\author{Hamed Merdji}
\affiliation{LIDYL, CEA, CNRS, Université Paris-Saclay,CEA Saclay 91191 Gif sur Yvette, France}

\author{Angel Rubio}
\affiliation{Max Planck Institute for the Structure and Dynamics of Matter, Luruper Chaussee 149, 22761 Hamburg, Germany.}
%\affiliation{Center for Free-Electron Laser Science CFEL, Deutsches Elektronen-Synchrotron DESY, Notkestraße 85, 22607 Hamburg, Germany.}
\affiliation{Nano-Bio Spectroscopy Group and ETSF, Departamento de Fisica de Materiales, Universidad del Pa\'is Vasco UPV/EHU, 20018, San Sebasti\'an, Spain.}
\affiliation{Center for Computational Quantum Physics (CCQ), The Flatiron Institute, 162 Fifth Avenue, New York, New York 10010, USA.}

\maketitle

\setcounter{figure}{0}

\makeatletter 
\renewcommand{\thefigure}{S\@arabic\c@figure}
\makeatother

%%%%%%%%%%%%%%%%%

{\bf Supplementary Note 1: Impact of the bandgap correction on HHG and IAP}  \\

\begin{figure*}
	\centering
	\includegraphics[width=1.\linewidth]{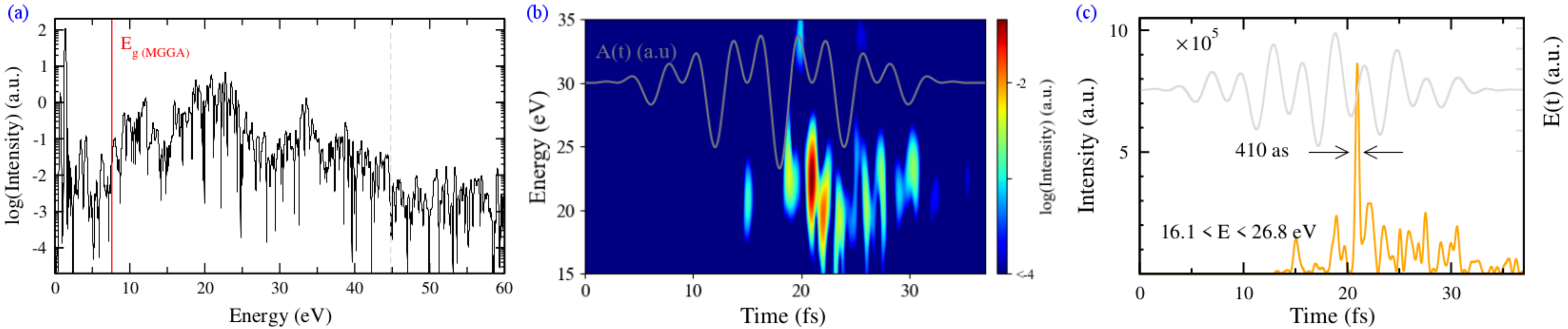}
	\caption{\label{mgga} {\bf The meta-GGA calculation results.}
		{\bf a} HHG spectrum corresponding to the pulse characteristics displayed in Fig.~\ref{hhg2color}a; the pulse polarization is along [100] direction. The dashed gray line marks the cutoff energy. The solid red line indicates the MgO bandgap in meta-GGA level which is equal to 7.6~eV.
		{\bf b} Time-frequency analysis of the HHG. The time window to calculate the Gabor transform is taken to be 0.25~fs. This figure also shows the time profile of the vector potential.
		{\bf c} The resulted attosecond pulse for the given energy window.}
\end{figure*}

\begin{figure*}
	\centering 
	\includegraphics[width=1.0\linewidth]{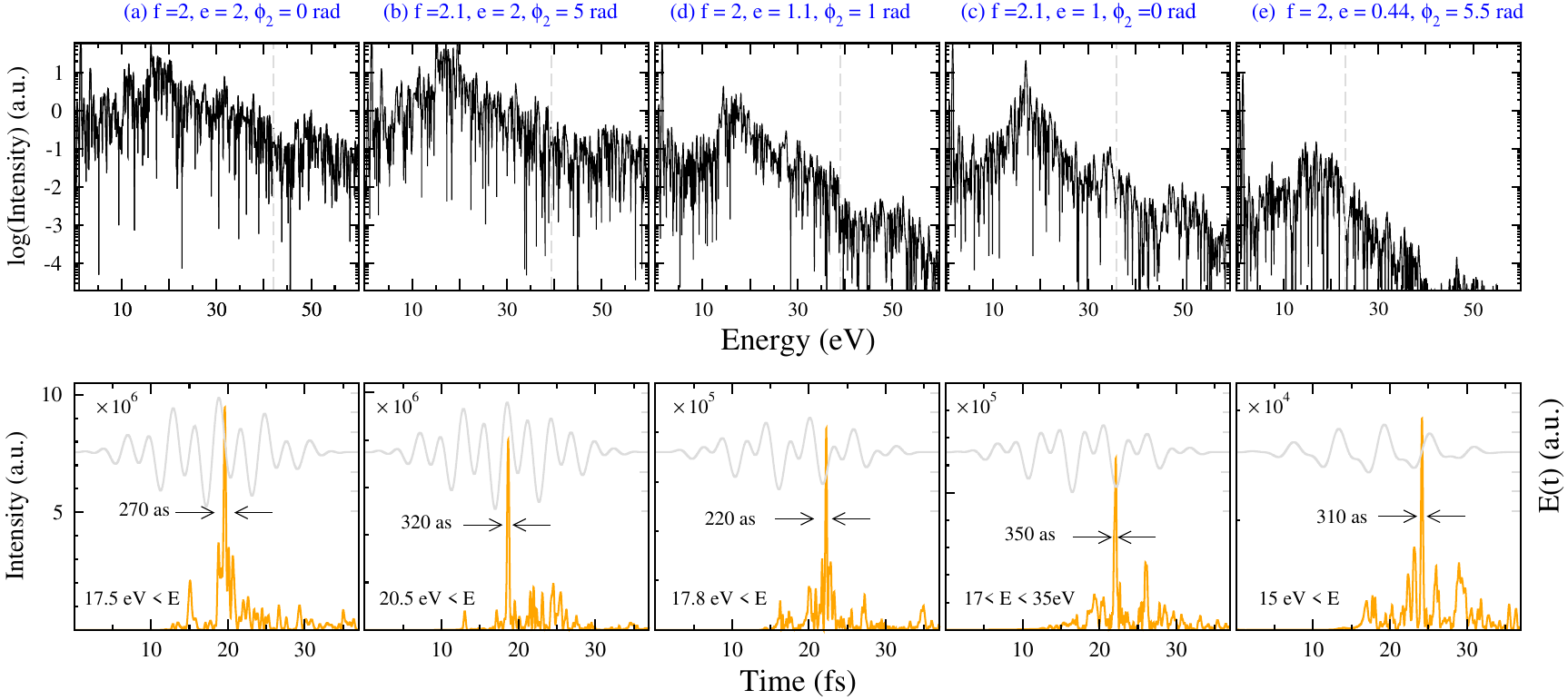}
	\caption{\label{hhg2color} {\bf Calculated HHG spectra (top row) and their related IAPs (bottom row) for different two-color pulses.} The titles specify the pulse characteristics; e, f and $\phi_2$ as defined in Eq.~1 in the main text. The light polarization is along [100] direction. The gray dashed lines in HHG spectra mark the cutoff energies.	The IAP duration and the corresponding energy windows are given in the bottom panels. The total electric field profiles are also shown in light gray in the bottom row, using the same scale for the panels.}
\end{figure*}

Since DFT calculations in LDA or GGA levels underestimate the bandgap; we employ the TB09 meta-GGA functionals, which yield an accurate estimation of MgO bandgap, to see the impact of the bandgap correction on HHG and IAP for the pulse characteristic displayed in Fig.~\ref{hhg2color}a. The results presented in Fig.~\ref{mgga} imply that the gap enhancement does not affect our main results; however, it weakens the HHG as well as IAP intensities for an order of magnitude, which can be easily understood in terms of a decreased ionization, as we opened the bandgap of MgO.\\

{\bf Supplementary Note 2: Impact of the pulse intensity on HHG and IAP}\\

\begin{figure*}
	\centering
	\includegraphics[width=0.4\linewidth]{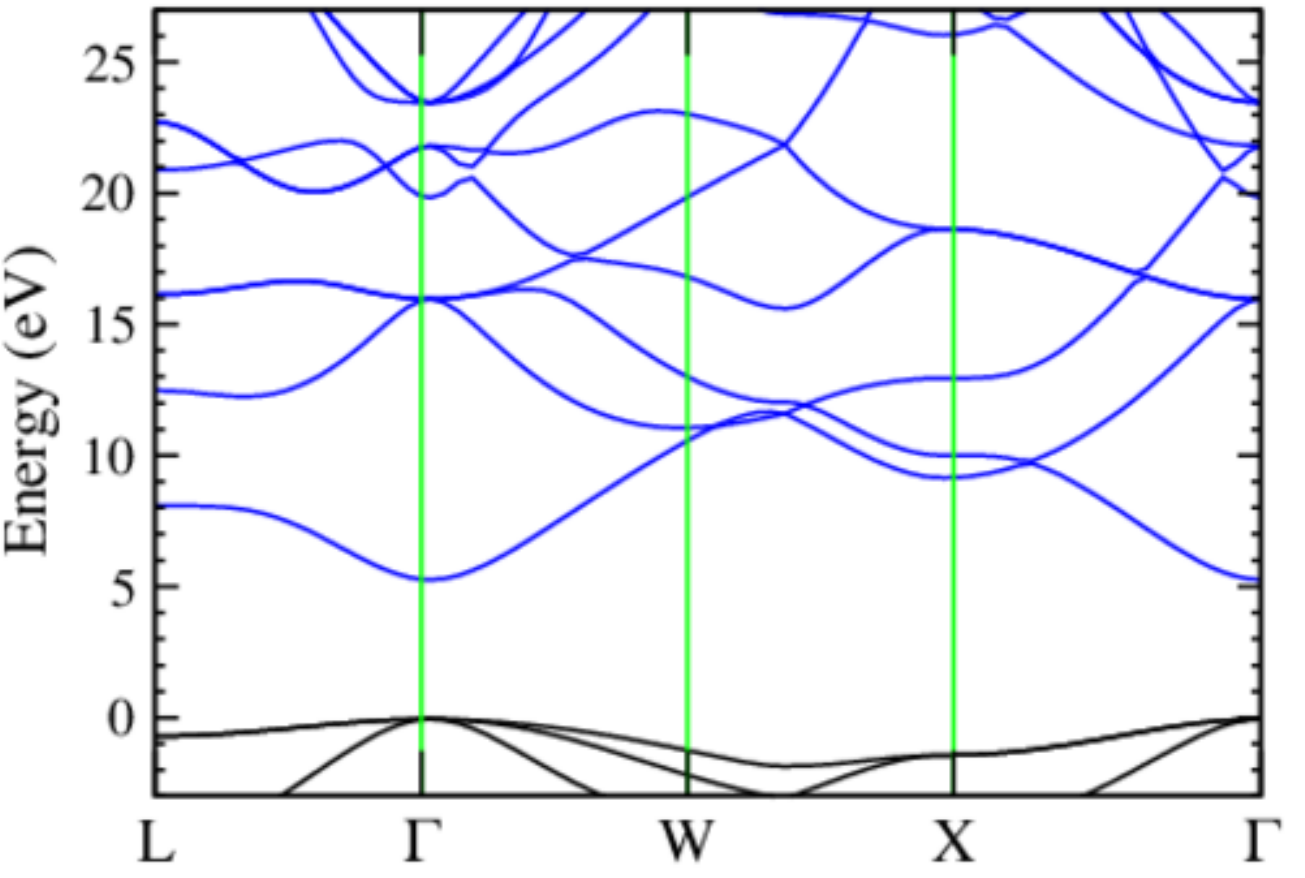}
	\caption{\label{attochirp} {\bf Bulk MgO DFT-LDA band structure.} It is shown in the fcc lattice high symmetry path; the valence (conduction) bands are in black (blue). The top of the valence band is set to be zero.} 
\end{figure*}

\begin{figure*}
	\centering
	\includegraphics[width=0.55\linewidth]{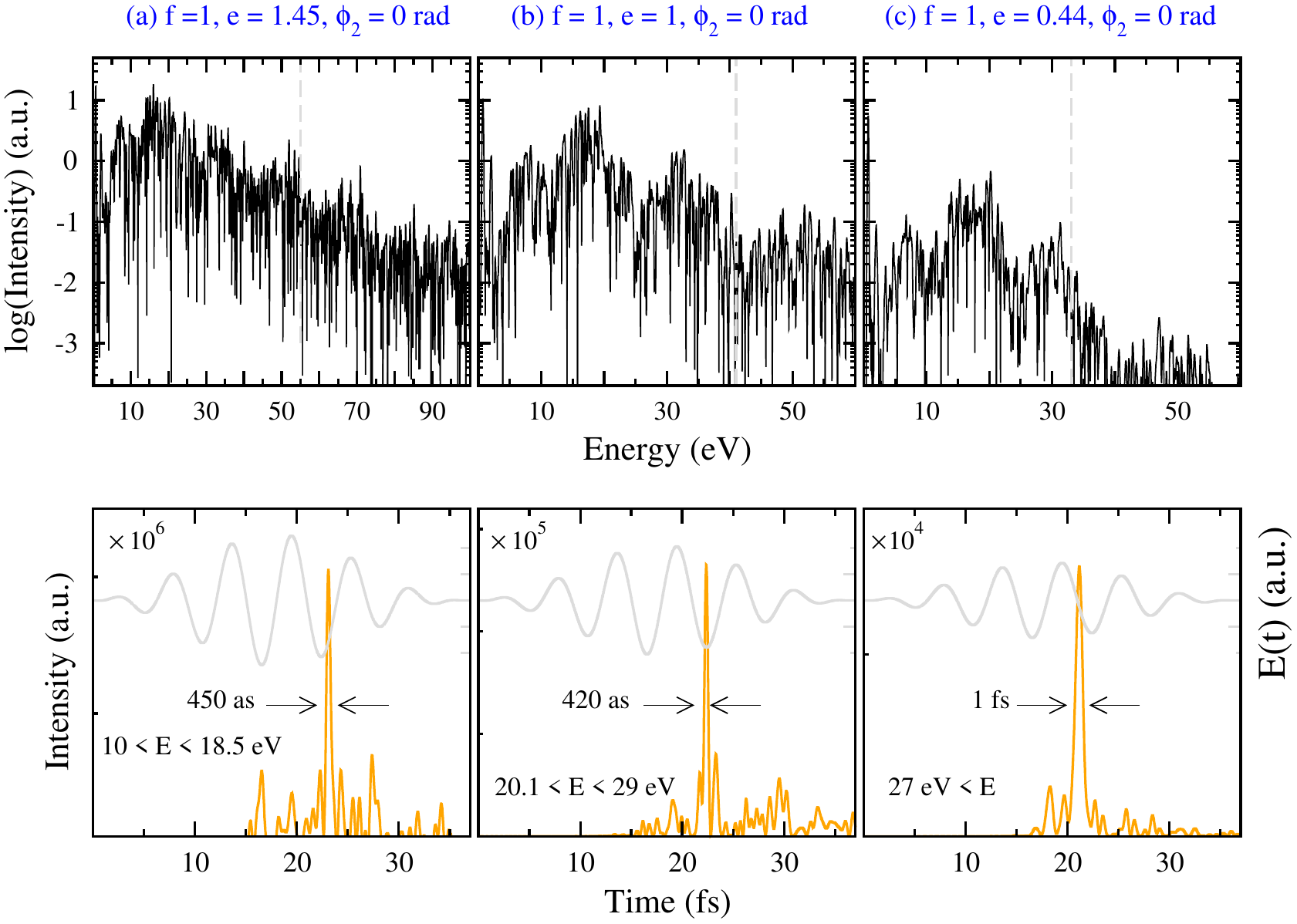} 
	\caption{\label{Ieff} {\bf Impact of the pulse intensity on high-harmonics and IAP generation for the case of one-color pulse.}   The pulse characteristics are same as the fundamental laser pulse, but from the right to the left, the pulse intensities are respectively $2I_0$, $4I_0$ and $6I_0$ where $I_0$ is the fundamental pulse intensity.}
\end{figure*}

The HHG spectra of the two-color pulses introduced in the main part are shown in Fig.~\ref{hhg2color}. The electric field increases from right to left.

It should be noted that the band structure has an important role to analyze the high-harmonics emission in solids;
because of the discrete band structure in solids, the cutoff energy definition for a HHG spectrum corresponding to a solid target is not as clean as that in gases. It is possible to define several cutoff energies in a HHG spectrum obtained from a solid.
We consider the  cutoff energy as the spectral position where the intensity decreases by an order of magnitude at a given harmonic and  beyond, averaged over a large bandwidth of several eV.

In order to evaluate the effect of driving pulse intensity in IAP generation,
Fig.~\ref{Ieff} presents one-color pulses HHG spectra. Here, the harmonic emissions are controlled just by the pulse intensity; the second pulse frequency is same as the fundamental pulse, and its phase is set to zero. 
In Fig.~\ref{Ieff} from left to right, the pulse intensities are respectively six, four, and two times the fundamental pulse intensity.  
This calculation clearly reveals that with an appropriate spectral filtering, IAPs can effectively be produced and optimized with the pulse intensity enhancement.
As shown in the main text, no IAP is generated from the fundamental pulse while the intensity enhancement makes it possible. This result is in contrast to what we concluded from two-color pulses. Since the intensity does not play any explicit role in the IAP, generated by the two-color pulse (shown in Fig.~\ref{hhg2color}); 
it reveals the strong impact and importance of asymmetric pulses in IAP generation. 
In addition, the two-color IAPs are much shorter and brighter than those achieved from the one-color laser. Regarding their wider energy window, they also show more flexibility compared to one-color-based IAP.

%%%%%%%%%%%%%%%%%%%%%%%%%%%%%%%%%%%%%%%%%

\begin{figure*}
	\centering
	\includegraphics[width=0.45\linewidth]{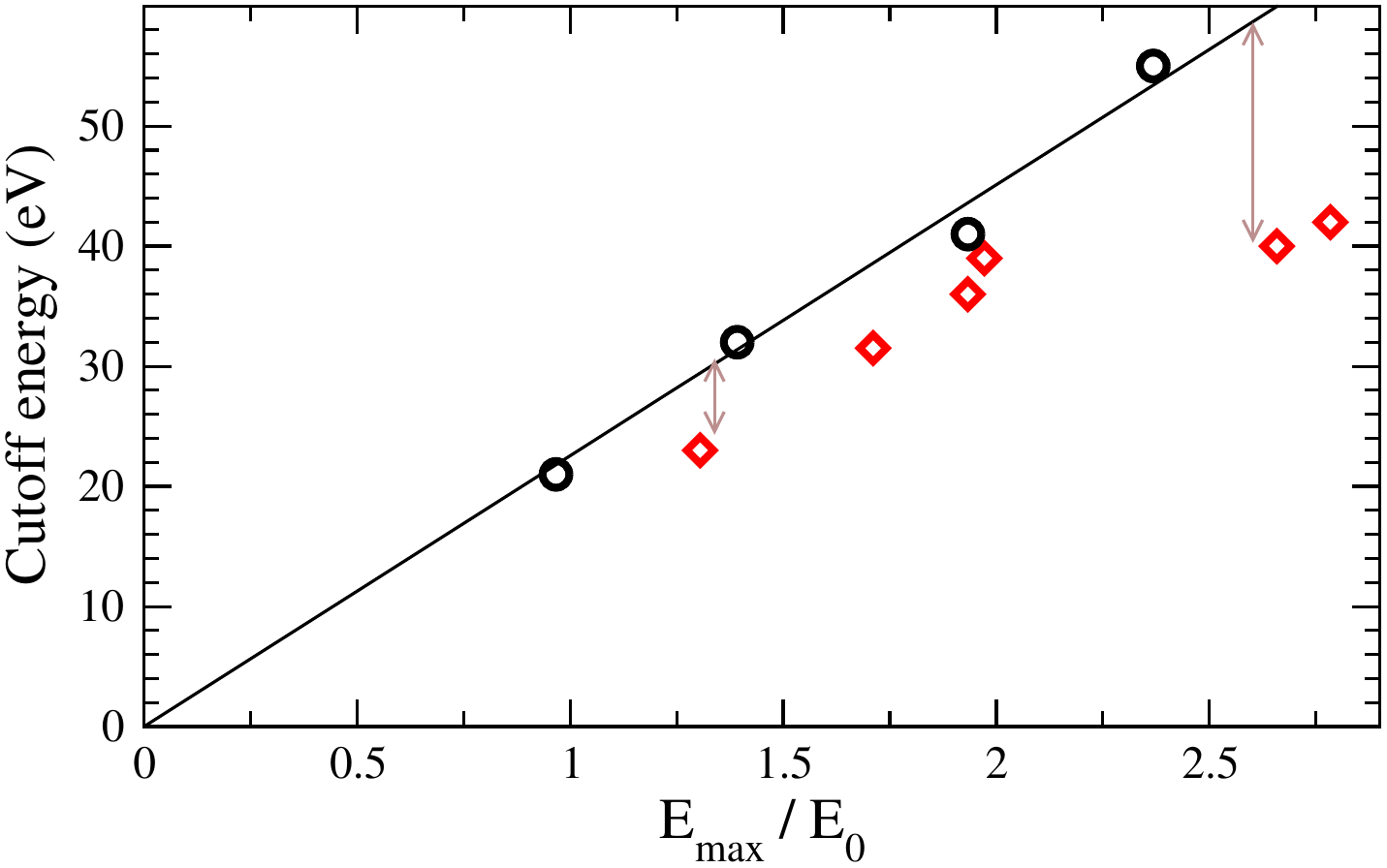} 
	\caption{\label{cutoff} {\bf Harmonic cutoff energies as a function of their electric field maximum.} The points are for the reported pulses in Figs.~\ref{hhg2color} and \ref{Ieff} as well as the fundamental pulse. E$_0$ and E$_\mathrm{max}$ denote the fundamental and the related pulse maximum electric field amounts, respectively. The black circles (red diamonds) relate to the one (two) color pulses. The brown arrows show that the same electric fields could lead to different cutoff energies which reveals the impact of the two color beating. Regarding to $e$ and $\phi_2$ parameters defined in the paper and Figs.~\ref{hhg2color} and \ref{Ieff}, if $\phi_2 = 0$, $E_\mathrm{max}/E_0 = 1+ e$. }
\end{figure*}

Figure~\ref{cutoff} shows the cutoff energy evolution as a function of the maximum value of the driving pulse electric field; this plot exhibits that the cutoff energy increases with the electric field enhancement. Regarding Fig.~\ref{cutoff}, the cutoff energies corresponding to the pulses with the same wavelength,  rise linearly as a function of the electric field strength.
But for the two-color pulses, with mixed wavelengths, the cutoff energies are not located on this line, and they have a lower value. 
Since our results predict that cutoff energy is not dependent to the wavelength (not shown), this shows the impact of non-symmetric pulse beating.\\

{\bf Supplementary Note 3: Impact of the ionic potential on HHG}\\  

The impact of the gradient of the electron-nuclei potential in HHG is discussed in Ref.~[17]; %\cite{hhgsolT-1}; 
it is shown that   
\be
HHG \propto \Big|FT\Big(\int_{\Omega}  d\mathbf{r}~n(\mathbf{r},t)\mathbf{\nabla} v_\mathrm{nuc}(\mathbf{r}) \Big) + N_e \mathrm{E}(\omega)\Big|^2\,,
\ee
where $v_\mathrm{nuc}(\mathbf{r})$ is the electron-ion potential, $N_e$ the number of electrons and $\mathrm{E}$ the driving electric field. Obviously the second term does not lead to non-perturbative harmonics, and thus the term $\mathbf{\nabla} v_\mathrm{nuc}(\mathbf{r})$ plays a key role in the harmonic emission. Since the [100] direction in MgO is along the Mg-O ionic bonds,  $\mathbf{\nabla} v_\mathrm{nuc}(\mathbf{r})$ in this direction has its largest value, and it is therefore expected to have a stronger emission for light polarization along [100] direction comparing to other directions.\\

%%%%%%%%%%%%%%%%%%%%%%%%%%%

{\bf Supplementary Note 4: Impact of the second pulse rotation on HHG}\\

\begin{figure*}
	\centering
	\includegraphics[width=1\linewidth]{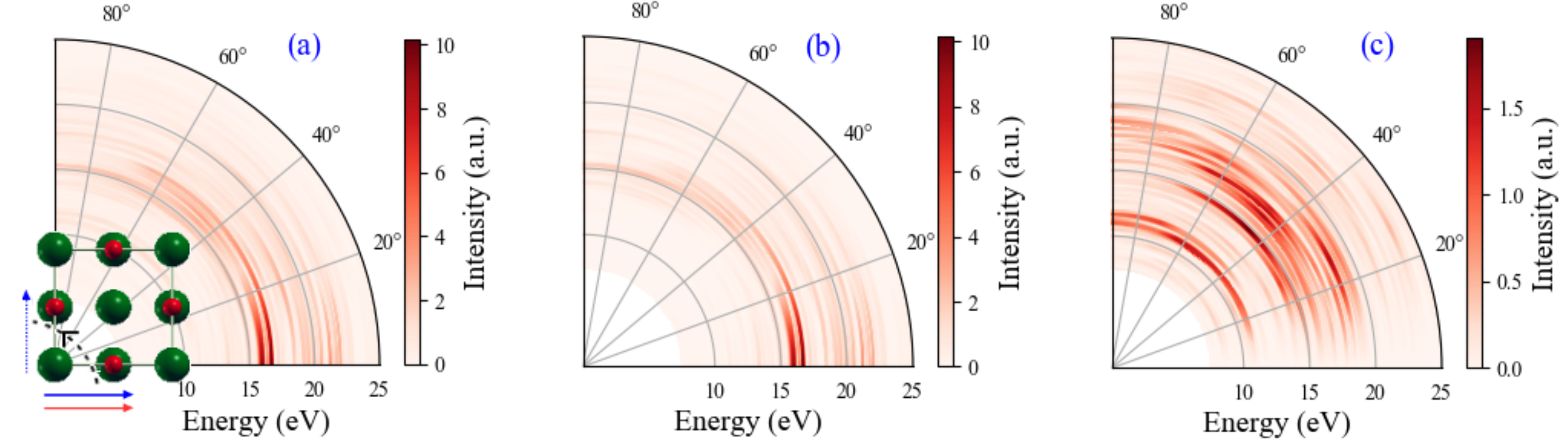}
	\caption{\label{polar} {\bf Two-color pulse HHG spectra as a function of the second field polarization direction.}  The contour displayed in {\bf a} shows the total HHG while {\bf b} and {\bf c} respectively indicate the high-harmonic spectra/emission with polarization along $x$ and $y$ directions. The first field polarization is fixed along the [100] direction. Note that the intensity scales are linear. The HHG with $x$ polarization, namely HHGx, is stronger than HHGy. }
\end{figure*}
\begin{figure*}
	\centering
	\includegraphics[width=0.5\linewidth]{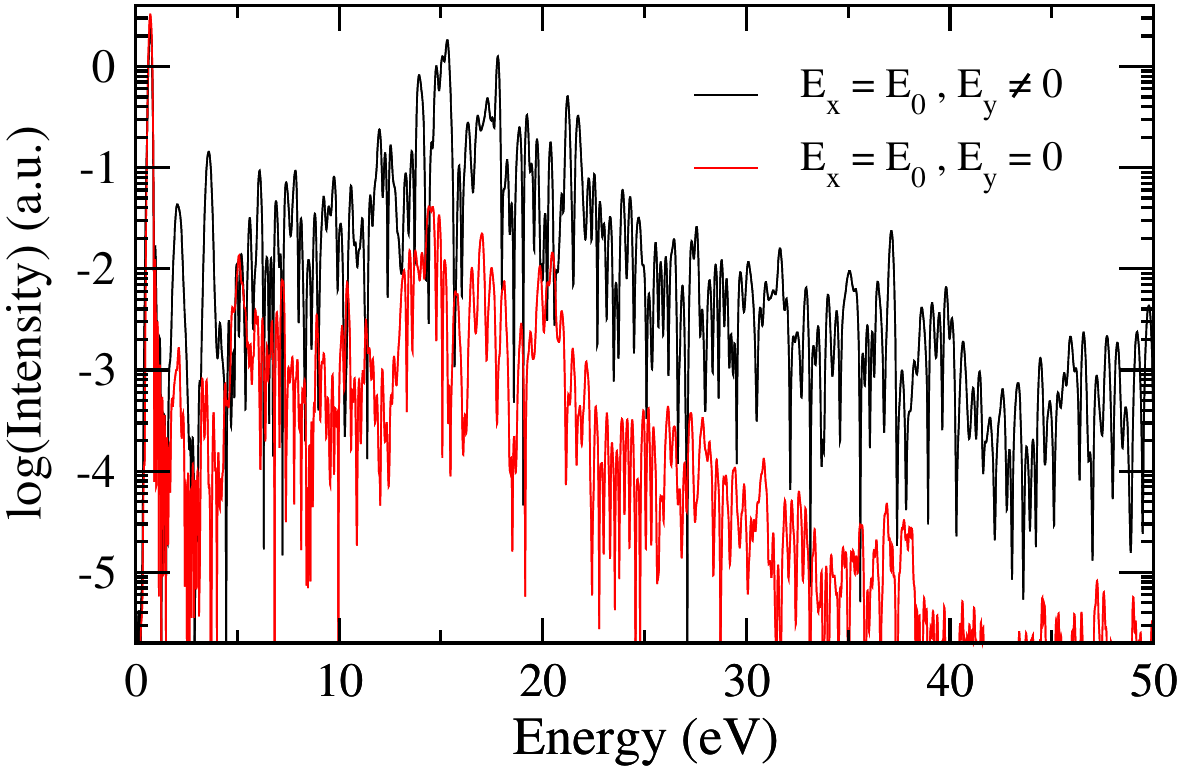} 
	\caption{\label{HHGx} {\bf Comparison between $x$ polarized HHG (HHGx) of two different pulses with the same electric fields in $x$-direction.} Red shows the fundamental pulse HHG (displayed in Fig.~1, too) and the black is the $x$-polarized HHG of the pulse displayed in Fig.~\ref{polar} when the second field polarization is along $y$ direction.}
\end{figure*}

Fig.~\ref{polar} displays the two-color HHG spectrum as a function of second pulse rotation angle  as well as the polarization resolved HHG spectra. The $x$-polarized HHG (HHGx) offers a declining trend \textit{versus} the second pulse rotation while the $y$-polarized HHG (HHGy) reaches its maximum around $\theta = 45^{\circ}$; it means that for  $45^{\circ} < \theta < 90^{\circ}$, HHGy is a decreasing function of $\mathbf{E}_y$ which implies the effect of $\mathbf{E}_x$ in $y$-polarized harmonics.
Figure~\ref{HHGx} gives another example of this behavior. This figure compares HHGx when the second applied pulse is rotated of 90$^\circ$  and the HHG resulting from the fundamental laser pulse only (as already shown in Fig.~1 in the paper). Since for both cases, the $x$ component of the total electric fields are equal, this figure reveals the role of the $E_y$ in enhancing the emission along the $x$ direction.
This behavior is compatible with what is expected from an elliptically polarized pulse in solids regarding the extended valence hole which strengthens the ellipticity response of solids;
additionally, it could make sense as the vertical electric field, here, the pulse along $y$ axis, could ionize the system and make the HHG process easier.

\end{document}